\documentclass[prc,twocolumn,floatfix,showpacs,superscriptaddress]{revtex4}
\usepackage{amssymb}
\usepackage[centertags]{amsmath}
\usepackage[paperwidth=210mm,paperheight=297mm,centering,hmargin=1.5 cm,vmargin=1.5 cm]{geometry}
\usepackage{txfonts}
\usepackage{bm}
\usepackage{epsfig}
\usepackage{graphicx,graphics}
\usepackage{float}

\begin{document}
\title{ Energy dependence of resonance production in relativistic heavy ion collisions }

\author{Feng-lan Shao}\email{shaofl@mail.sdu.edu.cn} 
\affiliation{School of Physics and Engineering, Qufu Normal University, Shandong 273165, China}

\author{Jun Song}\email{songjun2011@jnxy.edu.cn} 
\affiliation{Department of Physics, Jining University, Shandong 273155, China}

\author{Rui-qin Wang}
\affiliation{School of Physics and Engineering, Qufu Normal University, Shandong 273165, China}

\author{Mao-sheng Zhang }
\affiliation{School of Physics and Engineering, Qufu Normal University, Shandong 273165, China}

\begin{abstract}
	The production of hadronic resonances $K^{*0}(892)$, $\phi(1020)$, $\Sigma^{*}(1385)$, and $\Xi^{*}(1530)$ in central AA collisions at $\sqrt{s_{NN}}=$ 17.3, 200, and 2760 GeV are systematically studied. The direct production of these resonances at system hadronization are described by the quark combination model and the effects of hadron multiple-scattering stage are dealt with by a ultra-relativistic quantum molecular dynamics model (UrQMD). We study the contribution of these two production sources to final observation and compare the final spectra with the available experimental data. The $p_T$ spectra of $K^{*0}(892)$ calculated directly by quark combination model are explicitly higher than the data at low $p_T \lesssim 1.5$ GeV and taking into account the modification of rescattering effects the resulting final spectra well agree with the data at all three collision energies. The rescattering effect on $\phi(1020)$ production is weak and including it can slightly improve our description at low $p_T$ on the basis of overall agreement with the data. We also predict the $p_T$ spectra of $\Sigma^{*}(1385)$ and $\Xi^{*}(1530)$ to be tested by the future experimental data. 
\end{abstract}
\maketitle

\section{Introduction}
Collisions of heavy ions at extremely relativistic energies create the bulk matter with extremely high temperature and energy density \cite{QGP3}. The created matter expands, cools and finally produces thousands of hadrons of different species after hadronization. Comparing to the stable hadrons, the produced hadronic resonances is affected significantly by the subsequent hadron multiple-scattering process, because of their short lifetime (usually several fm/c) which is less than or comparable to the time span of the hadronic stage. Resonances such as $K^{*0}(892)$, $\phi(1020)$, $\Sigma^{*}(1385)$ have been measured by NA49 experiments at SPS energies, STAR experiments at RHIC energies and recent ALICE experiments at LHC energies, and rich experimental data of yields, transverse momentum $p_T$ spectra are reported \cite{DatKstarNA49,DatphiNA49,DatKstarSTAR,DatSigmaSTAR,datKstarPhiLHC}. Their production has been studied by several models or event generators such as UrQMD, AMPT and hybrid methods \cite{Bleicher02,Bleicher03,VogelAndBleicher05,VogelAndBleicher10,knospe2016,zkkstar2012,lkc2014}.

In this paper, we study the production of these hadronic resonance in relativistic heavy ion collisions by focusing on both the initial production dynamics at hadronization and the effects of hadronic rescattering stage. We choose three different collision system, i.e.,~central Pb+Pb collisions at $\sqrt{s_{NN}}=17.3$ GeV, central Au+Au collisions at 200 GeV and central Pb+Pb collisions at 2760 GeV, which covers two order of the magnitudes for collision energy. This can enable us to study the universal characteristic of resonance production in relativistic heavy ion collisions, taking advantage of the available data measured at SPS, RHIC and LHC. In particular, we discuss the initial production dynamics of hadronic resonances at hadronization. This is partially due to that the microscopic production dynamics of stable hadrons such as pions, kaons, protons, $\Lambda$, $\Xi$ and $\Omega$ has be well established by the experimental data. It is the quark combination mechanism that microscopically dominate their production, but not the traditional string fragmentation \raisebox{3pt}{\small{\footnotemark[1]}}.  Study of these resonance production will deep our understanding of the hadronization dynamics and test the existing hadronization models which have been tested against the experimental data of stable hadrons in relativistic heavy ion collisions. In order to present clearly our results, we adopt the two-step strategy. First, we use a combination model to give the $p_T$ spectra of resonance hadrons just after hadronization. Second, taking into the rescattering effect by a hadron transport model to give the final spectra of resonances and present them against the experimental data. We discuss whether the method of quark combination + hadron rescatterings can hold the essential characteristic of resonance production in relativistic heavy ion collisions or not.

\footnotetext[1]{Relativistic hydrodynamics and statistical model are also very popular for thermal hadron production in relativistic heavy ion collisions. They describe the macroscopic (thermodynamical) behavior of the transition of quark gluon matter to hadronic matter, which is not the microscopic dynamics of the formation of hadrons out of final state quarks and/or gluons we study here}

We mainly study four resonance hadrons, i.e.~$K^{*0}(892)$, $\Sigma^{*}(1385)$, $\phi(1020)$, and $\Xi^{*}(1530)$, which have different properties in their production. $K^{*0}(892)$ has the proper lifetime about 4 fm/c and thus will decay quickly once they are produced. Their decay products kaon and pion possibly scatter with the neighbor particles and thus lose the memory of their mother. On the other hand, pion and kaon has a certain probability to collide to regenerate the  $K^{*0}(892)$.
In addition, using available experimental data, we can study the production suppression of vector $K^{*0}(892)$ in quark combination model, relative to its low-lying partner kaon(497MeV). $\phi(1020)$ meson has a long lifetime about 46 fm/c, which is longer than the effective time duration of hadronic rescattering stage. The initially produced $\phi(1020)$ is hardly lost and $\phi(1020)$ can also be produced by two kaon coalescence and thus might has a rescattering effects to a certain extent. $\Sigma^{*}(1385)$ is baryon resonance, apart from the different reaction cross sections, the kinetics of daughter rescattering and regeneration is also different from $K^{*}$s. Thereby, we can study the baryon meson difference of hadronic rescattering effects.  $\Xi^{*0}(1530)$ is multi-strangeness hyperon, which can be used (together with $\Sigma^{*}(1385)$) to test the initial production of $J^{P}=(3/2)^+$ state relative to $J^{P}=(1/2)^+$ ground state and also the interaction dynamics of hyperon with light hadrons.

The paper is organized as follows. First we briefly introduce the quark combination model and hadronic rescattering model used in this work. Then we show the rescattering effects for the production of above resonances followed by a detailed discussion. Finally we show the final $p_T$ spectra of four resonances and compare them with available experimental data. The summary is given at last. 

\section{Brief introduction of working models}
As is well known, quark combination model provides a well and natural description of hadron production in the low and intermediate $p_T$ region for QGP hadronization \cite{Fries22003prl,Greco2003prl,hwa04,co2006PRC,sdqcm}, where the traditional string fragmentation fails. In this paper, we apply the quark combination model developed by Shandong group (SDQCM) \cite{sdqcm} to give the momentum spectra of hadronic resonances. Of all the ``on market' 'combination models, SDQCM is unique for its combination rule, which guarantees that mesons and baryons exhaust the probability of all the fates of the (anti-)quarks in a deconfined color-neutral system at hadronization. The main idea of the combination rule is to line up the(anti-) quarks in a one-dimensional order in phase space, e.g., in rapidity, and then let them combine into initial hadrons one by one according to this order \cite{sdqcm}. Three (anti-)quarks or a quark-antiquark pair in the neighborhood form a (anti-)baryon or a meson, respectively. The exclusive nature of the model makes it convenient to predict $K^{*0}(892)$ and other resonance  production on the basis of the reproduce of the yields and momentum spectra of various stable hadrons.  The model has been used to well explain the experimental data of yield, rapidity and $p_T$ spectra of various identified hadrons in relativistic heavy ion collisions at different energies \cite{sdqcm,wqr12,sjbbar13,wqr15charm, slx2012cpc,sj2009cpc}. In those previous works, we have studied production of stable hadrons including pions, kaons, protons, $\Lambda$, $\Xi$ and $\Omega$ at three collision energies. Assuming the same combination dynamics as those stable hadrons, we can directly compute the $p_T$ spectra of above resonance, using the obtained momentum distribution of quarks just before hadronization. 

\begin{figure}[htp]
	\centering
	\includegraphics[width=\linewidth]{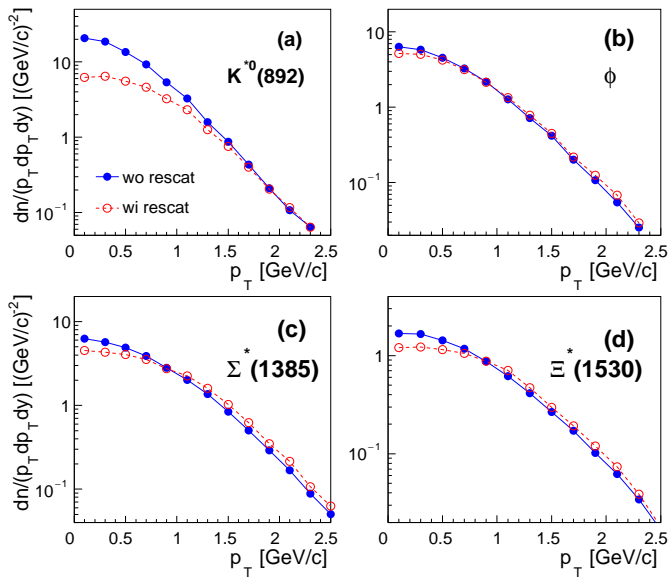}
	\caption{Mid-rapidity $p_T$ spectra of resonances with/without including effects hadronic rescattering stage in central Pb+Pb collisions at $\sqrt{s_{NN}}=17.3$ GeV.}
	\label{fig1}
\end{figure}

We use a ultra-relativistic quantum molecular dynamics model (UrQMD) \cite{urqmdRef,urqmdRef2} version 3.4 to calculate the effects of hadronic rescattering stage on resonance production.  The UrQMD model is a non-equilibrium transport approach and provides a well description of the phase-space evolution of hadron system. The interactions of hadrons include binary elastic and $2 \rightarrow n$ inelastic scatterings, resonance creations and decays, string excitations, particle-antiparticle annihilations as well as strangeness exchange reactions. The cross sections and branching ratios for the corresponding interactions are taken from available experimental measurements, detailed balance relations and the additive quark model. The model allows to study conveniently the full phase-space evolution of all hadrons and resonances in relativistic heavy-ion collisions. 
In this paper, we use the hybrid mode of the UrQMD and at the chemical freeze-out point we input our calculated resonance spectra for hadronization via Monte Carlo sampling method and then start the hadronic rescattering process. By comparing the spectra of initial resonance with those taking into account the effects of hadronic rescattering stage, we can study the presentation of two main production sources at final state. 

\section{Results of hadronic rescattering effects }
Fig.~\ref{fig1} shows the mid-rapidity $p_T$ spectra of $K^{*0}(892)$, $\phi(1020)$, $\Sigma^{*}(1385)$, and $\Xi^{*}(1530)$ directly produced by hadronization and those including effects of hadronic rescattering stage in central Pb+Pb collisions at $\sqrt{s_{NN}}=17.3$ GeV. Here, $\Sigma^{*}(1385)$ refers to $\Sigma^{*0}(1385) +\Sigma^{*-}(1385) $ plus their anti-particles in order to compare with the available experimental data and obtain high statistics.  $\Xi^{*}(1530)$ refers to $\Xi^{*0}(1530)+\Xi^{*-}(1530)$.
We see that hadronic rescattering stage causes an obvious suppression on low $p_T$ spectra of $K^{*0}(892)$ mainly because of the so-called signal loss, that is to say, $K^{*0}(892)$ decay products (pion and kaon) re-scattered by neighboring hadrons. With the increasing transverse momentum this signal loss become weak due to the prolonged fly lifetime (Lorentz effect). On the other hand, the regeneration of $K^{*0}(892)$ by pion kaon coalescence would contribute the yield density at intermediate $p_T$, as seen in $p_T \sim 2$ GeV in Fig.~\ref{fig1}(a). We see that the spectra of $\phi(1020)$ meson is slightly affected by hadronic rescatterings. For $\Sigma^{*}(1385)$ and $\Xi^{*}(1530)$, hadronic rescatterings also suppress their low $p_T$ spectra but the magnitude is smaller than that of $K^{*0}(892)$. And it causes a slight increase of the yield density at moderate $p_T$ region, which is the competition of signal loss and regeneration channels.

\begin{figure}[htp]
	\centering
	\includegraphics[width=\linewidth]{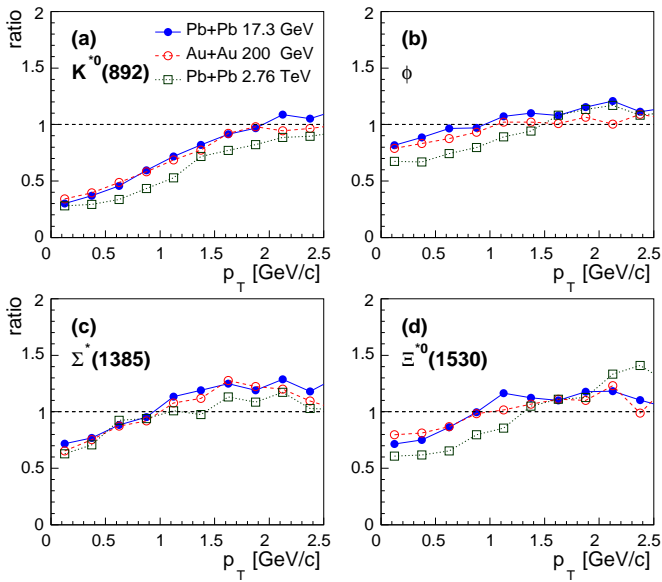}
	\caption{The ratio of the $p_T$ spectra of resonances incorporating  hadronic rescattering effects to those without at three different collision energies. The imperfect smoothness is due to the finite statistics.}
	\label{fig2}
\end{figure}

Effects of hadronic rescattering stage at different collision energies are shown in Fig.~\ref{fig2}. Here, in order to study their energy dependence, we plot the ratio of $p_T$ spectra of resonance incorporating the hadronic rescattering effects to those doesn't. Fig.~\ref{fig2}(a) shows results of $K^{*0}(892)$ in central AA collisions at $\sqrt{s_{NN}}=$17.3, 200 and 2760 GeV. We see that the production of $K^{*0}(892)$ is suppressed more than a half in the low $p_T$ and as $p_T$ increases the suppression becomes weak due to the contribution of regeneration channel. In addition, we see a weak energy dependence for the ratio of $K^{*0}(892)$ from 17.3 GeV to 200 GeV but an obviously continuous suppression as the collision energy increases to 2760 GeV, which is partially due to the much extended lifetime of hadronic phase at LHC energies. 
For $\phi(1020)$ in Fig.~\ref{fig2}(b), the ratio deviates from one around $20\%$ in covered $p_T$ region both at 17.3 GeV and 200 GeV, showing also a  weak energy dependence from SPS to RHIC energies. At LHC energy, the $\phi(1020)$ suppression at low $p_T$ region is relatively strong and reaches about 30\%. 
Comparing to panel (a), the hadronic rescattering effects of $\phi(1020)$ is obviously smaller than that of $K^{*0}(892)$.  This is closely related to $\phi(1020)$'s long lifetime and small reaction cross section between $\phi(1020)$ and other particle. Note that two kaon coalescence can regenerate $\phi(1020)$ and thus increase the $\phi(1020)$ yield at intermediate $p_T$ region to a certain extent (10-20\%)  which is determined by their cross section and their phase space populations in the system evolution. 
For $\Sigma^{*}(1385)$ and $\Xi^{*}(1530)$, we also find a nontrivial suppression at low $p_T$ region and a small increase at moderate $p_T$ due to the competition of signal loss and regeneration. 

\begin{figure}[!htp]
	\centering
	\includegraphics[width=\linewidth]{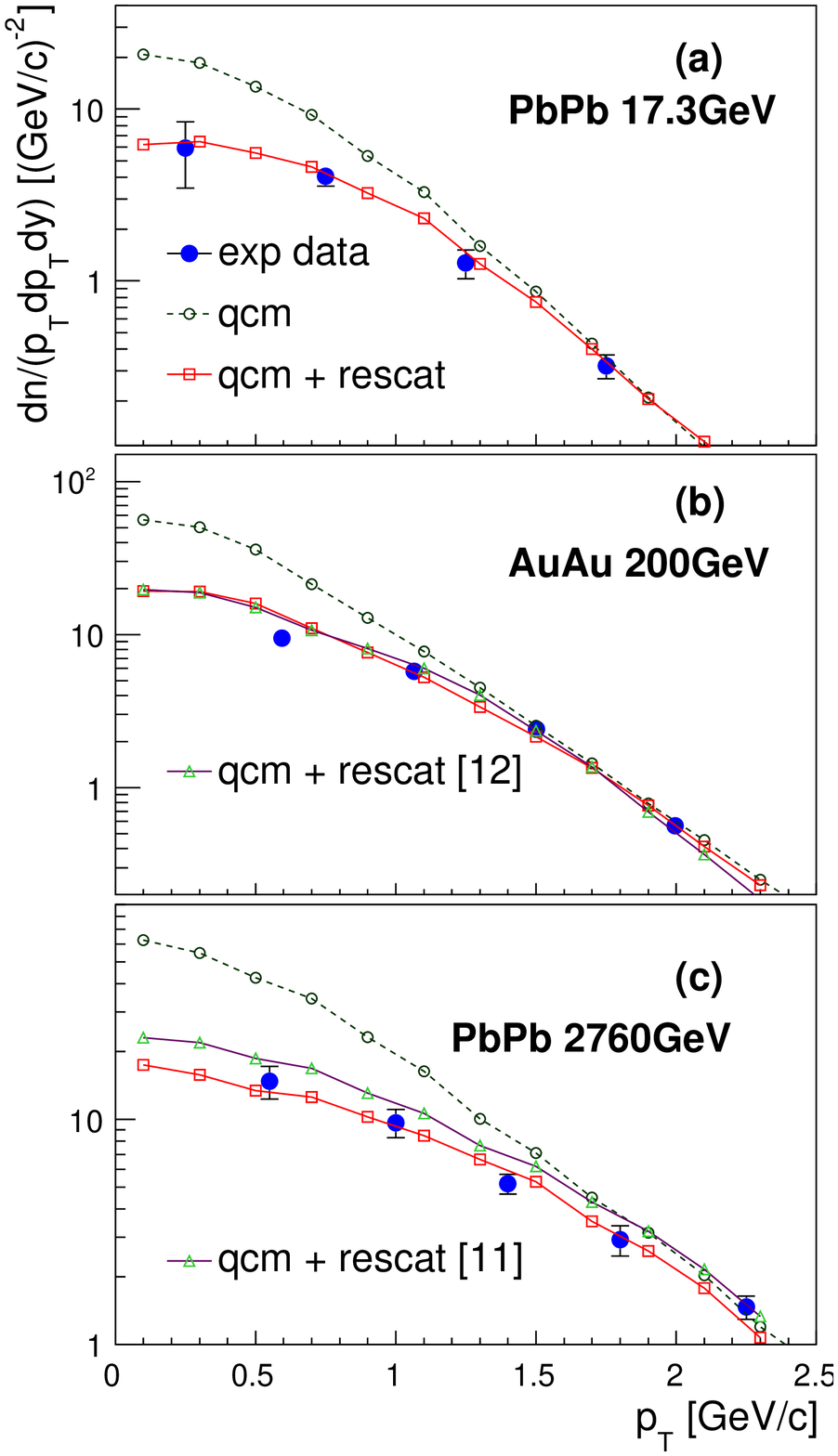}
	\caption{The $p_T$ spectra of $K^{*0}(892)$ at three different collision energies. Solid circles are experimental data \cite{DatKstarNA49,DatKstarSTAR,datKstarPhiLHC} and open symbols with lines are results of initially produced $K^{*0}(892)$ by SDQCM and those of taking into account of hadron rescattering effects, respectively. }
	\label{fig3}
\end{figure}

\begin{figure}[!htp]
	\centering
	\includegraphics[width=\linewidth]{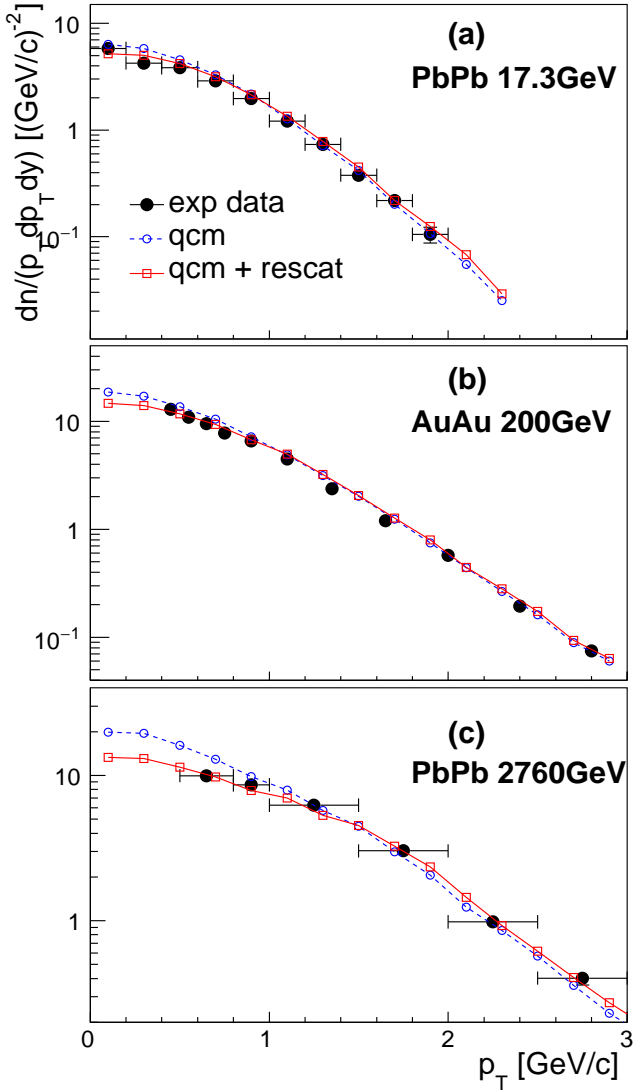}
	\caption{The $p_T$ spectra of $\phi(1020)$ at three different collision energies. Solid circles are experimental data \cite{DatphiNA49,DatPhiSTAR,datKstarPhiLHC} and open symbols with lines are results of initially produced $\phi(1020)$ by SDQCM and those of taking into account of hadron rescattering effects, respectively. }
	\label{fig4}
\end{figure}

\section{Final spectra of resonances}
In Fig.~\ref{fig3}, we present the $p_T$ spectra of initial $K^{*0}(892)$ and final ones taking into account of rescattering effects at three different collision energies, and compare them with the experimental data \cite{DatKstarNA49,DatKstarSTAR,datKstarPhiLHC}. The open circles with dashed lines are $p_T$ spectra of direct $K^{*0}(892)$ given by SDQCM. We see that at low $p_T$ region $p_T \lesssim 1$ GeV, $K^{*0}(892)$ produced by hadronization is obviously greater than the data at three collision energies. As $p_T$ increases, QCM results begin to close to the data. Note that these direct spectra of $K^{*0}(892)$ is calculated using the momentum distribution of constituent light quarks and strange quarks obtained in previous works where we have well reproduced the experimental data of various stable hadrons, in particular, that of kaons. The over-estimation at low $p_T$ show the necessity of including the significant effects of hadronic rescattering stage. Taking into these effects, we get the final spectra of $K^{*0}(892)$, which are shown as the open squares with solid lines in Fig. \ref{fig3} (a-c).  In panel (a) for Pb+Pb 17.3 GeV, we obtain a well description of experimental data. For panel (b) at Au+Au 200 GeV, we also get an improved agreement with the data. Here, we also plot another result, up-triangles with solid line, in which the hadronic rescattering effect is calculated from the ART subroutine in AMPT model obtained in our previous work \cite{zkkstar2012}. We find that it is very close to the present one. At LHC energy in panel (c), consideration of the hadronic rescattering effects significantly suppresses the low $p_T$ $K^{*0}(892)$ production and the final $p_T$ spectrum tends to agree  with the experimental data of ALICE collaboration \cite{datKstarPhiLHC}. Here, we also plot another result in which the hadronic rescattering effect is calculated in Ref.~\cite{knospe2016}, and it can be regarded as a reference of the theoretical uncertainty of hadronic stage.  The obtained final $K^{*0}(892)$ spectrum is shown as the up-triangle with solid line in panel (c). This result exhibits weaker suppression than that we calculated. It slightly exceeds the data in $p_T\lesssim 2$ GeV about 20\% but agrees with the last data point. Together these two results, we argue that the experimental data can be explained by the method of this paper. The results of Fig.~\ref{fig3} thus gives us the significant confidence that the initial production dynamics of $K^{*0}(892)$ at hadronization should also be the quark combination. This is philosophically satisfactory that hadrons with the same flavor constituent have the same production mechanism. 

In meson formation in quark combination model, the relative production probability of the lowest-lying vector meson (V) to pseudo-scalar meson(P) with the same constituent quark composition is tuned by the parameter $R_{V/P}$. In previous work \cite{zkkstar2012} at high RHIC energies, we found the value of about 0.45 for $R_{V/P}$ can explain the data of kaons and $K^{*0}(892)$ simultaneously. Here, we see that using the same value the data of both SPS and LHC energies are also well explained. This universal parameter value suggests a nontrivial constrains for modeling the spin-induced interaction dynamics at hadronization in the more sophisticated quark combination model in future.

Fig.~\ref{fig4} shows the results of $\phi(1020)$ meson and the comparison with the experimental data \cite{DatphiNA49,DatPhiSTAR,datKstarPhiLHC}. As shown in Fig.~\ref{fig2}, the rescattering effects of $\phi(1020)$ production is weak, which is due to its long lifetime and small interaction cross with other particles and relatively rare two kaon coalescence. The final $\phi(1020)$ $p_T$ spectra including the rescattering effects, open-squares with solid line, change little relative to the initial ones given by SDQCM, open-circles with dashed lines. We find a well agreement with the experimental data at SPS (panel (a)) and RHIC (panel (b)). As the collisional energy increases to LHC, see panel (c), the rescattering effects suppress, to a certain extent, the yield density of the $\phi(1020)$ meson at low $p_T$($\lesssim 1$ GeV) and we observe an visible decrease for the final $\phi(1020)$ spectra there, leading to an improved agreement between our final result and the experimental data of ALICE collaboration.

\begin{figure}[!htp]
	\centering
	\includegraphics[width=\linewidth]{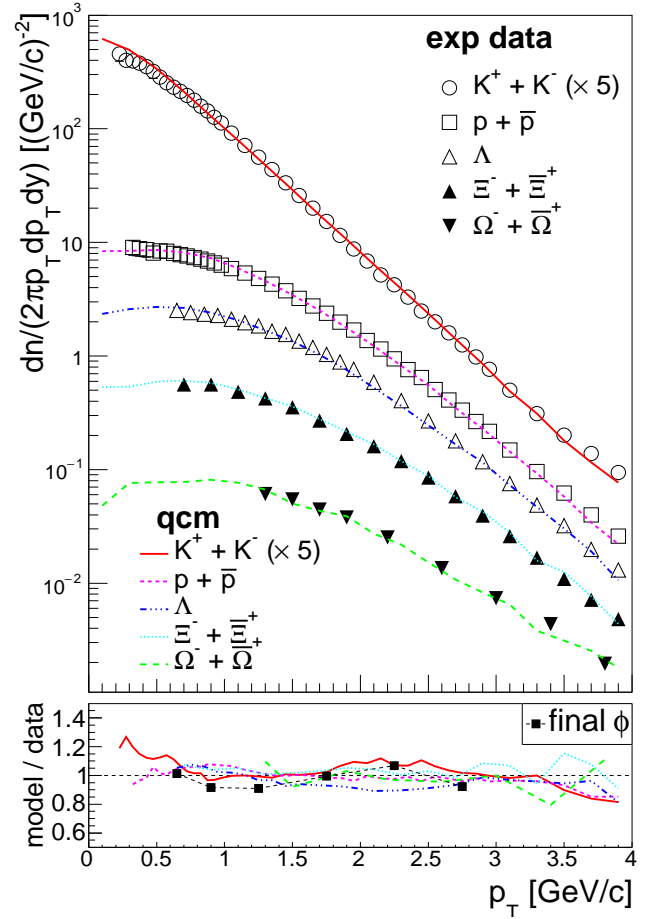}
	\caption{The $p_T$ spectra of stable hadrons in central Pb+Pb collisions at $\sqrt{s_{NN}}=2760$ GeV.  Symbols are the experimental data \cite{DatProtonLHC,DatKaonProtonLHC,DatXiOmegLHC,DatLamLHC} and lines are results of SDQCM. The data and model result of kaon is multiplied by the factor 5 for clarity.}
	\label{fig5}
\end{figure}

\begin{figure}[!htp]
	\centering
	\includegraphics[width=0.9\linewidth]{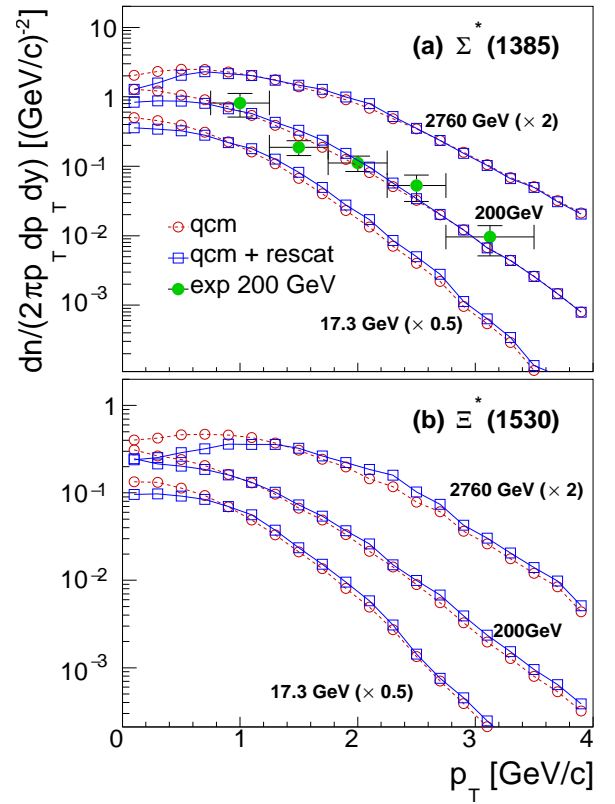}
	\caption{The $p_T$ spectra of $\phi(1020)$ with/without incorporation of hadronic rescattering effect at three different collision energies.}
	\label{fig6}
\end{figure}
Some comments on $\phi(1020)$ results at LHC energies are in order. The data show the $p_T$ spectrum of proton is very close to that of $\phi(1020)$ \cite{datKstarPhiLHC}, which is expected in hydrodynamical model mainly due to their similar mass. So it is argued in Ref.~\cite{datKstarPhiLHC} that the $\phi(1020)$ meson production is thermal dominated at LHC energies. Surely, thermalization is more important at LHC. However, we note that the shape of $\phi(1020)$ distribution calculated by hydrodynamical model still deviate the data to a certain extent \cite{datKstarPhiLHC,vishSong11}, i.e. it overestimates the production about 50\% at $p_T \lesssim 2$ GeV but underestimates the data almost the same proportion at larger $p_T$.  In addition, hydrodynamical model fails to reproduce the $\Omega/\phi(1020)$ ratio as the function of $p_T$ as $p_T \gtrsim 2$ GeV \cite{datKstarPhiLHC}. Note that the exponential behavior of the hadronic $p_T$ spectra at LHC extend to 4 GeV for meson and 6 GeV for baryons due to the large collective flow \cite{DatKaonProtonLHC}. On the other hand, at LHC QCM still perform as well as that at RHIC energies, as shown in Fig.~\ref{fig4} for $\phi(1020)$ and our previous work \cite{wqr15charm} for stable hadrons. To illustrate it more clearly, Fig. \ref{fig5} shows the $p_T$ spectra of various stable strange hadrons obtain in our work \cite{wqr15charm} with further refined quark spectra for thermal components. The $p_T$ region is chosen $0-4$ GeV, because we observe that the $\Omega/\phi(1020)$ ratio linearly increase about to 4 GeV \cite{datKstarPhiLHC} showing the dominated region for thermal quark combination. The relative difference between our results with the data is less than about 20\%, which is smaller than those of hydrodynamical model and others reported in literatures \cite{datKstarPhiLHC, DatXiOmegLHC,vishSong11,HKMv1,HKMv2,KARKOW1,KARKOW2}. 

Finally, we shows the results of final $\Xi^{*}(1530)$ and $\Sigma^{*}(1385)$ in Fig.~\ref{fig6}.  At present, only the data of $\Sigma^{*}(1385)$ at Au+Au collisions are available and are shown also in panel (a). We see that our results at 200 GeV are basically agreement with the data. The results of at SPS and LHC are left to be tested by the future experimental data. 
For $\Xi^{*}(1530)$, the rescattering effects at LHC are stronger than that for $\Sigma^{*}(1385)$ but are still smaller than that for $K^{*0}(892)$. We predict their final spectra for the future test by experimental data. 

\section{Summary}
	We have studied the production of hadronic resonances $K^{*0}(892)$, $\phi(1020)$, $\Sigma^{*}(1385)$, and $\Xi^{*}(1530)$ in relativistic heavy ion collisions, using the hybrid method of quark combination hadronization plus hadronic multiple scattering effects. The experimental data of  $K^{*0}(892)$ and $\phi(1020)$ at central Pb+Pb collisions at  $\sqrt{s_{NN}}=$ 17.3, central Au+Au collisions at 200 GeV and central Pb+Pb collisions at 2760 GeV are well explained. We find that the effects of hadronic rescattering causes the significant suppression on $K^{*0}(892)$ production at low $p_T$ but slight suppression for $\phi(1020)$ production. The effects of hadronic rescattering is also nontrivial for $\Sigma^{*}(1385)$, and $\Xi^{*}(1530)$ production. We predict the final $p_T$ spectra of $\Sigma^{*}(1385)$ and $\Xi^{*}(1530)$ for the future test by experiments.

\section{Acknowledgments} The work is supported by the National Natural Science Foundation of China under Grants No. 11575100 and No. 11305076.

\end{document}